# Photocatalytic hydrogen production of Co(OH)$_2$ nanoparticle-coated α-Fe$_2$O$_3$ nanorings

Heberton Wender,*[a] Renato V. Gonçalves,[b] Carlos Sato B. Dias,[ac]
Maximiliano J. M. Zapata,[b] Luiz F. Zagonel,[c] Edielma C. Mendonça,[ad]
Sérgio R. Teixeira[b] and Flávio Garcia[a]

The production of hydrogen from water using only a catalyst and solar energy is one of the most challenging and promising outlets for the generation of clean and renewable energy. Semiconductor photocatalysts for solar hydrogen production by water photolysis must employ stable, non-toxic, abundant and inexpensive visible-light absorbers capable of harvesting light photons with adequate potential to reduce water. Here, we show that a-Fe$_2$O$_3$ can meet these requirements by means of using hydrothermally prepared nanorings. These iron oxide nanoring photocatalysts proved capable of producing hydrogen efficiently without application of an external bias. In addition, Co(OH)$_2$ nanoparticles were shown to be efficient co-catalysts on the nanoring surface by improving the efficiency of hydrogen generation. Both nanoparticle-coated and uncoated nanorings displayed superior photocatalytic activity for hydrogen evolution when compared with TiO$_2$ nanoparticles, showing themselves to be promising materials for water-splitting using only solar light.

## 1 Introduction

Due to their environmentally benign nature, and depletion of the world's fossil-fuel reserves, hydrogen-based energy systems are increasingly attracting attention worldwide.[1] A potential route for clean energy generation is the use of solar power to efficiently split water into H$_2$ and O$_2$ molecules. This process, known as photocatalytic water splitting, was first reported by A. Fujishima and K. Honda in the early 1970s, using TiO$_2$ as a semiconducting material.[2] The process is based on photon absorption events generated by electron-hole pairs in a semiconductor structure where photo-generated electrons reduce water to form H$_2$ while holes oxidize water to form O$_2$.[3] If a voltage bias is applied between the semiconductor and any counter electrode, the process is called photoelectrocatalysis (PEC).[4]

However, the simplest and the most economical way to split water is by using sunlight as the only energy source. This process is considered an artificial photosynthesis, called pho- tocatalysis (PC), where a powder photocatalyst dispersed in water or aqueous mixture is illuminated by an external light source. Even though use of photon energy conversion using powdered photocatalysts is not yet industrially viable, consid- erable advancements in this direction have been made.

In order to have good properties for H$_2$ evolution, photo- catalysts must display suitable conduction and valence band levels as well as appropriate band gap widths. The bottom level of the conduction band has to be more negative than the proton reduction potential (H+/H$_2$ = 0 V vs. the normal hydrogen electrode (NHE)), while the top level of the valence band needs to be more positive than the oxidation potential of water (1.23 eV). The absorption of photons is only limited by the band gap.[3-7] This band structure is a thermodynamic requirement but not a sufficient condition in itself. Different processes need to take place in order to achieve an effective water splitting: (i) photon absorption—generation of electrons and holes with sufficient potentials; (ii) charge separation—migration to surface reaction sites; (iii) suppression of recombination between electron-hole pairs, and (iv) construction of surface reaction sites for both H$_2$ and O$_2$ evolution.[8]

[a]*Brazilian Synchrotron National Laboratory (LNLS), CNPEM, Rua Giuseppe Máximo Scolfaro 10.000, Postal Code 6192, 13083-970, Campinas, SP, Brazil. E-mail: hbtwender@gmail.com*

[b]*Laboratório de Filmes Finos e Fabricação de Nanoestruturas (L3Fnano) UFRGS, Instituto de Física, Av. Bento Gonçalves 9500, P.O. Box 15051, 91501-970, Porto Alegre, RS, Brazil*

[c]*Instituto de Física Gleb Wataghin, Universidade Estadual de Campinas-UNICAMP, Rua Sergio Buarque de Holanda 777, 13083-859 Campinas, SP, Brazil* [d]*Universidade Federal de Sergipe (UFS), Itabaiana, SE, Brazil*

The published version has also electronic supplementary information (ESI): Histograms of size distributions of NP-coated and as-prepared IONRs (Fig. S1 and S2); selected-area EDS spectrum of the IONR/Co(OH)$_2$ NPs (Fig. S3); STEM images with details of the IONR/NP surface (Fig. S4); diffuse reflectance data (Fig. S5) and XPS survey spectra of IONRs (Fig. S6). See DOI: 10.1039/c3nr02195e

In this context, different materials have been tested as catalysts for water splitting. Most of them are derived from wide band gap metal oxides such as $TiO_2$,[9,10] and $Ta_2O_5$,[11] which are active only under ultra-violet (UV) excitation (~5% of solar energy power). In parallel, layered Ti-based perovskites,[12] titanates,[13] and tantalates,[3] among many others, have also been studied as alternatives. Extension of the absorption window to the visible range of the solar spectrum has been successfully performed by doping[10] (electron donor level formation), band gap narrowing[7] (solid-solution reaction), and sensitization with organic dyes,[14] either with or without sacrificial agents. Another important strategy is to use direct visible light-driven photocatalyst(s) such as GaP,[15] InP,[16] $Fe_2O_3$,[4] oxynitrides,[17,18] metal sulfides[19,20] and oxysulfides.[21]

Hematite (a-$Fe_2O_3$), a semiconductor oxide of band gap 2.1 eV, derived from the fourth most common element in the earth's crust, has emerged as a promising alternative, due to its chemical stability in water, abundance, low cost and significant light absorption.[4] The main problem with using hematite for water splitting is that its conduction band level is below the redox potential of $H+/H_2$.[3] Nevertheless, different strategies have been recently proposed to overcome this problem, such as: (i) reducing to nanoscale size (with control over shape)[22] - to enhance the number of surface-active sites and to reduce the bulk recombi- nation of electron-hole pairs at the same time -;[23-26] (ii) doping - to mitigate problems caused by energy mismatch between water redox potentials and the band edges of hematite -;[27] (iii) ion irradiation - to decrease the resistivity and increase the donor density and flatband potential[28] - and (iv) surface modification with other oxides - to achieve appropriate band-edge characteristics.[29] Following from this, a direct substantial band gap increase compared to bulk hematite was revealed in low-dimen- sional nanomaterials,[30] raising the possibility of using hematite nanomaterials directly, without external bias, for efficient hydrogen generation through water photolysis.

In the present study, iron oxide nanorings (IONRs) were synthesized through a hydrothermal reaction and coated with different concentrations of cobalt hydroxide nanoparticles (NPs). These nanomaterials were applied as photocatalysts to study their role in hydrogen evolution by water photolysis. These nanorings were specially chosen to investigate the effects of nano sizing and ring shape in the photo-response of pure hematite. Furthermore, the efficacy of using $Co(OH)_2$ NPs as co-catalysts was investigated, with a view to improving the hydrogen evolution. The materials were characterized by using different advanced techniques, in order to obtain a complete understanding of the results. In this way, the optimum amount of $Co(OH)_2$ NPs was found, and some interesting points regarding the mechanisms for hydrogen generation are discussed in detail, including hydrogen production with pure IONRs.

## 2 Experimental section
### 2.1 Synthesis and coating of the iron oxide nanorings (IONRs)

Hematite (a-$Fe_2O_3$) NRs were prepared following the same procedure reported by Jia *et al.*[31] In a typical experiment, $FeCl_3$ (0.02 M), $NaH_2PO_4$ (0.18 mM) and $Na_2SO_4$ (0.55 mM) were mixed and magnetically stirred for 10 min at room temperature. A total volume of 80 mL was transferred to a Teflon-lined autoclave reactor of 110 mL capacity, that was closed and maintained at 220 °C for 48 h. After cooling to room temperature, a red powder could be obtained as a precipitate. This was washed three times with ethanol to remove possible residues, centrifuged, and dried at 80 °C under vacuum conditions.

Cobalt hydroxide nanoparticles were deposited on the surface of the previously prepared IONRs following chemical precipitation. In a typical procedure, 2.5 mL of a 0.1 M solution of $CoCl_2$ was mixed with 20 mL of distilled water containing 0.01 M of the previously prepared a-$Fe_2O_3$ NRs. After mixing, the solution was heated to 60 °C under magnetic stirring, and 40 mL of NaOH (0.01 M) was added drop by drop. The final solution was kept at 60 °C for 30 min and cooled to room temperature. The samples were obtained by centrifugation and dried at 60 °C. A fraction of the sample was kept as-synthesized while the other one was annealed at 300 °C in air.

### 2.2 Characterization

The morphology of the samples was investigated using a FEI Inspect F50 Field Emission Scanning Electron Microscope (FESEM), operated at 10 and 30 kV with a secondary electron detector. A High-Resolution Transmission Electron Micro- scope (HRTEM, model JEOL JEM 3010) operated at 300 kV was used to investigate both the morphological and crystal- line features of the IONRs and NPs. Electron Energy Loss Spectroscopy (EELS) was performed using a JEOL 2100F equipped with a Field Emission Gun (FEG) operating at 200 kV with an energy resolution of about 1 eV. EELS was obtained using a Gatan GIF Tridiem. The data were acquired in Scanning Transmission Electron Microscopy (STEM) mode in the form of spectrum images (the electron beam is focused on the sample and a spectrum is acquired for each position, forming a tri-dimensional dataset). The spectrum images were de-noised *via* the principal component analysis method using Hyperspy, a free software for hyperspectral data anal- ysis.[32,33] The spectra were calibrated using the hematite NRs as standard and using the spectra from A. Gloter *et al.* as reference.[34]

X-ray diffraction (XRD) patterns were recorded using a Phi- lips X'PERT diffractometer with Cu Ka radiation (l = 1.54 A) at 2θ — 20-90° with a 0.02° step size by measuring 5 s per step. X- ray absorption near edge spectroscopy (XANES) experiments were conducted at the XAFS1 and XAFS2 beamlines of the Bra- zilian Synchrotron Light Laboratory (LNLS).

### 2.3 Photocatalytic reactions

The photocatalytic reactions were carried out in a double-walled quartz photochemical reactor.[35] The temperature of the reaction system was maintained at a constant 25 °C by circulating water through the quartz photochemical reactor with a thermostatic bath. All the experiments were performed taking 4.0 mg of pho- tocatalyst powder suspended by magnetic stirring in 8 mL of ethanol-water solution (23.8 vol%), with a pH of 7. A 240 W Hg-Xe lamp (PerkinElmer; Cermax-PE300) was used as a light source. The quartz photochemical reactor was positioned at a distance of 20 cm of lamp housing. Prior to irradiation, the system was deaerated by bubbling argon for 30 min, followed by vacuum to remove any other gases inside the reactor. The gases produced by water photolysis were quantified using gas chromatography at room temperature in an Agilent 6820 GC chromatograph equipped with a thermal conductivity detector (TCD) with a Porapak Q (80/100 mesh) column by using argon as the carrier gas. The amounts of gases produced were measured at intervals of 0.5 h using a gas-tight syringe with a maximum volume of 100 mL. To check the reproducibility of the photocatalytic activity, each sample was measured three times.

## 1 Results

### 3.1 Iron oxide nanoring (IONR) synthesis

Fig. 1 presents a FESEM image of the as-prepared NRs. It shows that IONRs of well-defined structure could be obtained by a conventional hydrothermal reaction with an aqueous solution of $FeCl_3$ in the presence of sulfate and phosphate ions as additives.[31] The IONRs displayed mean inner diameter, outer diameter and height of about 37, 96 and 72 nm respectively, as can be seen in the ESI, Fig. S1.f The nanorings showed a rela- tively low polydispersity, mainly for outer diameter (12%) and height (10%).

The oxidation state and crystallinity of the as-synthesized IONRs were investigated by measuring XANES at the Fe K edge and XRD, respectively (Fig. 2). The XANES spectrum of the IONRs completely agreed with the standard a-$Fe_2O_3$ (Sigma- Aldrich) measured under the same conditions, which showed that the iron oxidation state is purely +3. Moreover, the XRD pattern could be attributed to the corundum structure of a-$Fe_2O_3$ (PDF number 33-0664).

Rietveld refinements show that the refined cell parameters are $a$ — 3.06 and $c$ — 13.76 and that the grain size of the a-$Fe_2O_3$ is about 129.3 nm with a preferred orientation in the [104] direction. Both results support the formation of hematite single phase in the as-prepared IONRs. The formation mechanism of this interesting nanostructure has already been described[31] and is in agreement with the results obtained here.

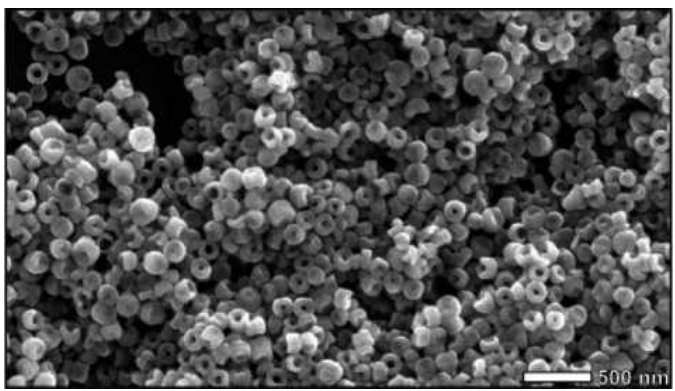

**Fig. 1** FESEM image of the iron oxide nanorings prepared by the hydrothermal reaction.

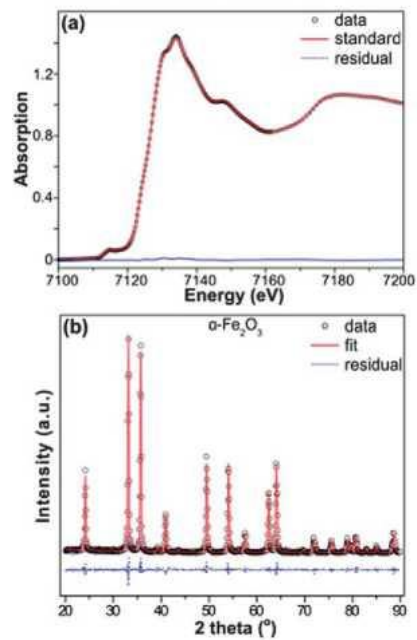

**Fig. 2** Normalized IONRs and standard a-$Fe_2O_3$ (Sigma-Aldrich) XANES spectrum (a) and XRD pattern with Rietveld refinement (b) of the hydrothermally prepared IONRs. Both results confirmed that the as-prepared nanorings are composed of pure hematite phase.

### 3.2 Cobalt oxide nanoparticle-coated IONRs

The previously prepared IONRs were used as starting materials for a second process in which cobalt oxide NPs were chemically deposited on their surface. Fig. 3a shows the XRD pattern of the product obtained after the chemical precipitation process using a 100 mM $CoCl_2$ solution as a starting reagent (see the Experimental section for details). The diffraction pattern was indexed to two crystalline phases, one composed of hematite (PDF 33-0664) and another corresponding to $Co(OH)_2$ (PDF 30-443). The data were refined by using the Rietveld method in order to quantify the amount of hydroxide deposited on the IONR surface as well as the crystal size of both phases. These samples were also prepared by using different molar concentrations of $CoCl_2$ ranging from 5.75 to 100 mM.

Rietveld refinement results are summarized in Table 1, and show that $Co(OH)_2$ was in the form of small NPs (less than 30 nm). The relative amount of $Co(OH)_2$ NPs increased from 5.3% to 45.0% when the initial concentration of $CoCl_2$ increased from 12.5 to 100 mM (see Table 1). With respect to the $Co(OH)_2$ NP crystal size, it was shown to be approximately constant at 13 nm for $CoCl_2$ concentrations ranging from 5.75 to 50 mM. However, the crystal size increased to 29.4 nm in the case of 100 mM of $CoCl_2$ in the initial reaction. When the $CoCl_2$ initial concen- tration was 5.75 mM, the cobalt based NPs could not be detected by X-ray diffraction due to being below the detection limit. In this case, a linear regression was used to estimate the amount of $Co(OH)_2$ NPs present in the sample (see Table 1).

In order to transform the cobalt hydroxide to cobalt(II, III) oxide, one selected sample was subjected to a process of annealing at 300 °C for 4 h. After this process, the $Co(OH)_2$ could be completely transformed to $Co_3O_4$, as observed by XRD (Fig. 3b). Here, the relative amount of Co$_3$O$_4$ is 20% with a crystalline size of about 7 nm. It is important to point out that the crystalline size of cobalt NPs was reduced after the annealing process what might be due to phase transformation from Co(OH)$_2$ to Co$_3$O$_4$.

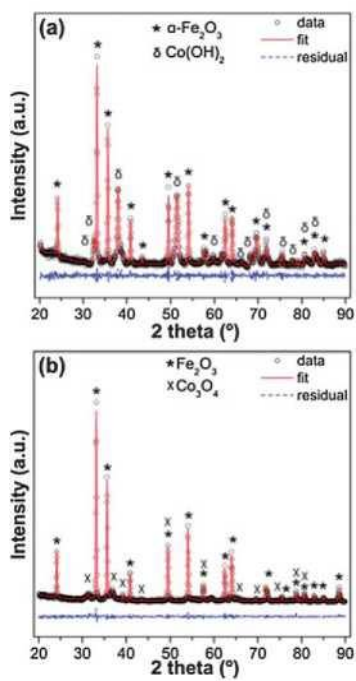

**Fig. 3** XRD pattern and Rietveld refinements of the cobalt oxide-coated IONRs: (a) as-prepared and (b) after annealing at 300 °C for 3 h.

**Table 1** Concentration and crystalline size of Co(OH)$_2$ and Fe$_2$O$_3$ NRs measured using Rietveld refinement

| CoCl$_2$ conc. (mM) | Co(OH)$_2$ conc. (%) | Co(OH)$_2$ Cs[b] (nm) | IONR Cs (nm) |
| --- | --- | --- | --- |
| 5.75 | Not detected[a] | | 130.5 ± 0.8 |
| 12.5 | 5.3 ± 0.5 | 13.60 ± 0.04 | 131.0 ± 0.6 |
| 25 | 7.0 ± 0.4 | 12.80 ± 0.06 | 144.4 ± 0.5 |
| 50 | 11.0 ± 0.9 | 12.90 ± 0.04 | 129.5 ± 0.6 |
| 100 | 45.0 ± 0.8 | 29.40 ± 0.04 | 146.4 ± 0.7 |

[a] Estimated to be 2.3 ± 0.6 by fitting a linear relation to the other points. [b] Cs means "crystal size".

To investigate the morphology of the NPs as well as the IONRs, Secondary electron images were taken using a FESEM microscope (Fig. 4). It was observed that the morphology of the IONRs was not strongly affected by the chemical precipitation of the Co(OH)$_2$ NPs on their surface. Essentially, the mean size increased after the reaction, as shown in Fig. 4a and S2 (ESI).f Moreover, two different regions could be observed in the FESEM images; one containing the IONRs coated with NPs homogeneously distributed on their surfaces (Fig. 4b); and other regions containing agglomerates of Co(OH)$_2$ NPs, evidenced by EDS in the respective area, Fig. S3 of ESI.f These agglomerated regions can be better visualized in the STEM images of Fig. S4 of ESIf, as well as in Fig. 4c (red arrow).

No significant changes were observed in either morphology or mean size of the NRs after the annealing process (Fig. 4c and S4 of ESIf). Furthermore, it is clear that the NRs were coated with Co(OH)$_2$ NPs on their surface (Fig. 4b) which were transformed to Co$_3$O$_4$ by annealing (Fig. 3b).

The NR-NP interface characteristics were further investigated by HRTEM and EELS. Fig. 5a shows a HRTEM image of the IONRs after the chemical precipitation of Co(OH)$_2$ NPs on their surface. It confirmed that the NPs adhere to the IONR surface, in some cases forming a semi-spherical shape. The darker region in Fig. 5a corresponds to the IONRs where an interplanar distance of 3.68 Å assigned to (012) planes of hematite could be seen. Two well-defined interplanar distances could be observed in the FFT image taken from an oriented nanoparticle (Fig. 5b), namely 2.34 Å and 2.93 Å. These distances match to a good approximation with (101) and (100) planes of Co(OH)$_2$, respectively. Fig. 5c shows a FFT image taken in the IONR region indicating its single-crystalline form.

EELS analysis revealed that the NPs are made only of cobalt oxide (no iron signal could be detected within the NPs for the acquisition parameters used). Fig. 5d and e show chemical maps in the region of some NPs. These maps were reconstructed from acquired spectrum images. By careful inspection of the chemical maps, it is possible to extract spectra from the NPs without any contribution from the nearby iron oxide NRs. Typical spectra are shown in Fig. 5f and g. The position of the cobalt L$_3$ edge peak of the as-prepared sample (after chemical precipitation of Co) shifts to higher energies after annealing, which is in agreement with the transition from Co$^{2+}$ to Co$^{3+}$. The annealed NPs also display an oxygen K-edge peak that matches the energy and shape of the Co$_3$O$_4$, while the spectrum from the sample without annealing shows more details in the fine structure, compatible with what should be expected from cobalt hydroxide.[36,37]

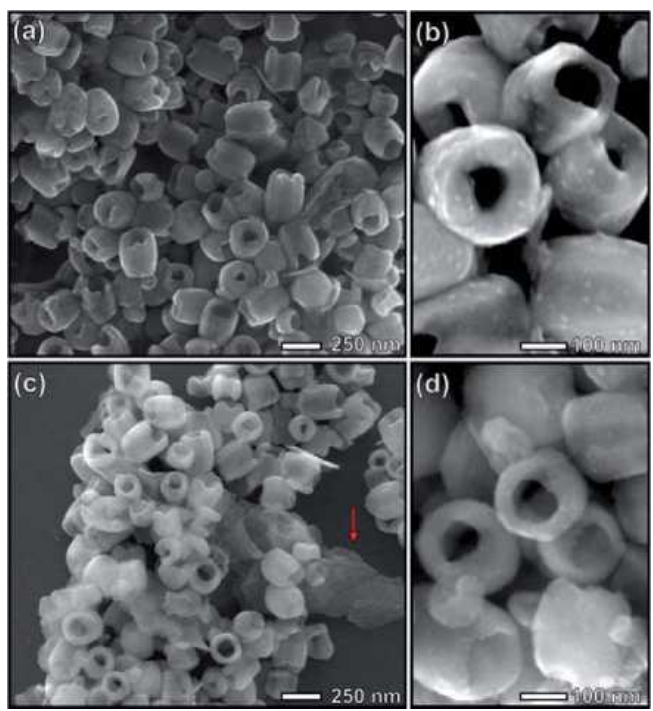

**Fig. 4** FESEM images of the NP-coated hematite NRs: as prepared samples (a and b) and after annealing up to 300 °C for 4h in an air atmosphere (c and d). It is possible to see that the NR surface is coated with NPs.

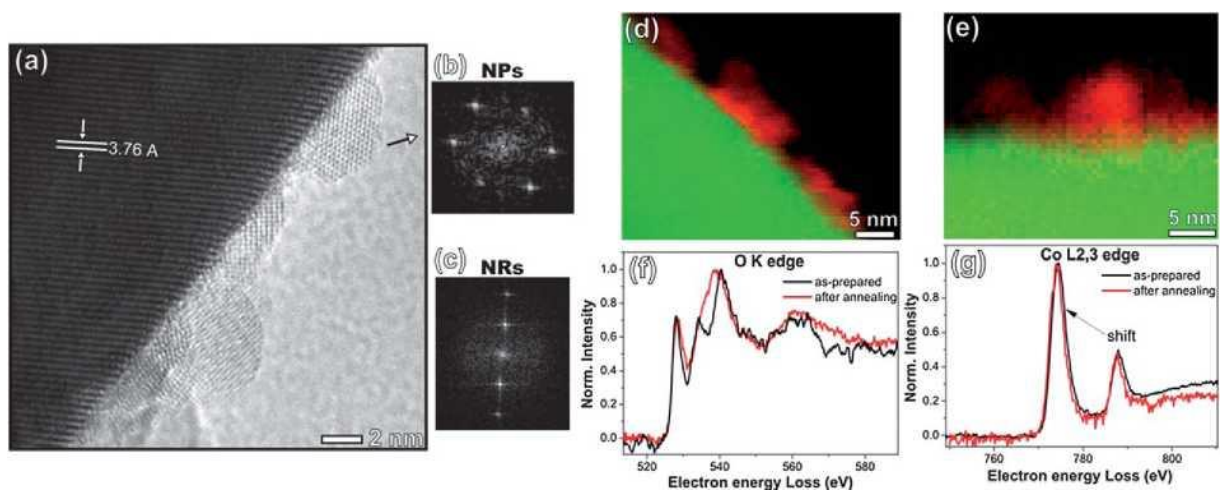

**Fig. 5** H RTEM image of the as-prepared Co(OH)$_2$ NP-coated IONRs (a), FFT image of the NPs (b), and of the darker region corresponding to the IONRs (c), chemical mapping obtained by EELS wherein the green color corresponds to the NRs and the red color to the NPs before (d) and after annealing (e), O K edge (f), and Co L2,3 edge (g) EELS spectra obtained by scanning a region containing only the NPs.

### 3.3 Photocatalytic properties

The Fe$_2$O$_3$ NRs, as well as the Co(OH)2 NP-impregnated NRs, were applied as photocatalysts for hydrogen production by photolysis. In order to find its optimal concentration, the pho- tocatalytic activity for H$_2$ production was evaluated using Fe$_2$O$_3$ NRs containing different amounts of Co(OH)$_2$ NPs. Fig. 6a shows the H$_2$ evolution rates for the different concentrations studied. When the pure Fe$_2$O$_3$ NRs were used as photocatalysts a H$_2$ evolution rate of about 350 mmol h$^{-1}$ g$^{-1}$ could be observed. However, in the presence of small amounts of Co(OH)$_2$ NPs (from ~2.3% to 5.3% of Co(OH)$_2$) on the NR surface, the photocatalytic activity increased to approximately 420 mmol h$^{-1}$ g$^{-1}$. A maximum H$_2$ evolution of about 546 mmol h$^{-1}$ g$^{-1}$ was obtained with 7% of Co(OH)$_2$ NPs. After this point, for concentrations of 11.0% and 45.0%, the photocatalytic activity decreased. From these results, it was possible to infer that the optimum amount of Co(OH)$_2$ NPs on the NR surface is near 7%.

The photocatalytic performance for hydrogen generation was also investigated after thermal treatment at 300 $^0$C for 4 h to investigate and compare the efficiency of Co(OH)$_2$ and Co$_3$O$_4$. Both samples were compared with commercial TiO$_2$ NPs (P25- Degussa) under the same experimental conditions (Fig. 6b). The results show that the photocatalytic activity of Fe$_2$O$_3$/Co(OH)$_2$ is higher than that of Fe$_2$O$_3$/Co$_3$O$_4$ and TiO$_2$ (P25), with values of 546, 392 and 140 mmol h$^{-1}$ g$^{-1}$ respectively.

### 3.4 Discussion

The results presented herein indicate that the Co(OH)2 NP-coated and pure Fe$_2$O$_3$ NRs are suitable materials for hydrogen generation by photolysis of water under UV-visible light excitation. The photocatalytic activities of the NR/NP composites were greater than commercial P25 NPs measured under the same experimental conditions.

It is noteworthy that bulk a-Fe$_2$O$_3$ has a suitable band gap of about 2.1 eV for water splitting but possesses a conduction band edge at an energy level below the reversible hydrogen potential.[4] The band gaps of Fe$_2$O$_3$ NRs and Co(OH)2 NP-impregnated NRs were investigated using diffuse UV-vis reflectance spectroscopy, Fig. S5 of ESI.f These results revealed a band gap of 2.16 eV, very close to that reported for bulk hematite, and that it was not altered by chemical precipitation of the Co(OH)$_2$ NPs on the NR surface.

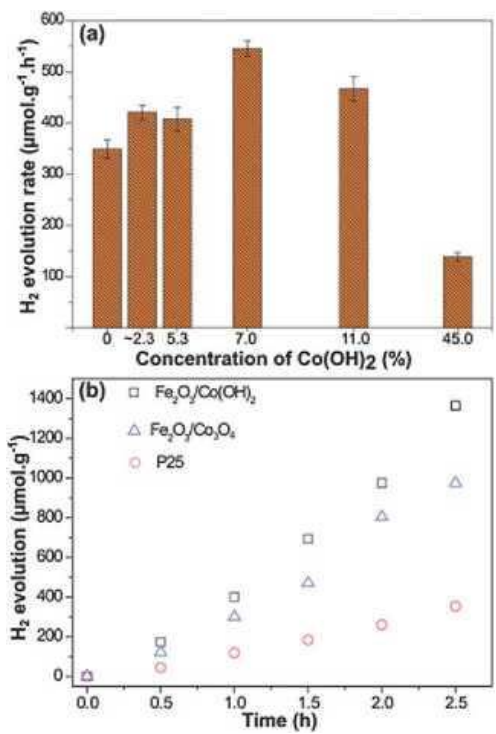

**Fig. 6** (a) Hydrogen evolution rate of Fe$_2$O$_3$ NRs impregnated with different concentrations of Co(OH)$_2$ NPs. The concentration of ~2.3% was estimated as presented in Table 1. (b) Comparison of the H$_2$ photocatalytic production of Fe$_2$O$_3$/Co(OH)$_2$, Fe$_2$O$_3$/Co$_3$O$_4$, and TiO$_2$ NPs (P25).

In addition, it was reported that hematite has a small diffusion length and consequently a high rate of electron-hole recombination.[4] These aspects play a major role in limiting the application of iron oxide as a photocatalyst for water splitting reactions. However, the results presented herein show that even for pure hematite NRs the photocatalytic activity was signifi- cantly higher than, for example, $TiO_2$ NPs under the same experimental conditions. In this case, engineering hematite to the nanoring shape probably changed the band edge positions for a region suitable for hydrogen generation using ethanol as a sacrificial agent at pH 7, as also evidenced before.[30] It still needs to be experimentally investigated by using advanced techniques; this will be the subject of forthcoming studies.

In addition, no contamination could be seen in the NR surface as observed by XPS (Fig. S6, ESIf), which reinforces the idea that the shape and size could increase the photocatalytic performance. These findings suggest that the photocatalytic activity observed for $H_2$ generation through water splitting is probably due to improved light absorption and high surface area and reactant transport.[30,38] For comparison, bulk hematite was subjected to the same photolysis conditions and no detectable photocatalytic activity could be seen.

In addition, coating with appropriate amounts of $Co(OH)_2$ NPs through chemical precipitation improved the photo- catalytic activity of the $Fe_2O_3$ NRs in $H_2$ production. This suggests that the $Co(OH)_2$ NPs present on the NR surface are operating as co-catalysts, which act as electron traps for the electrons migrating to the NR surface, thereby preventing recombination of electrons and holes. It probably enhanced the photocatalytic activity by providing reaction sites at the NR surface, and also increased the lifetime of electrons.[39,40] The increase in the mean $H_2$ evolution rate reached about 35%, when compared to pure hematite NRs, at the optimum amount of $Co(OH)_2$ NPs (near 7%, as determined by Rietveld) as co-catalysts. Moreover, the photoactivity decreased significantly (about 75%) when the amount of $Co(OH)_2$ NP co-catalyst exceeded the optimum range. This may be due to the blocking of the semiconductor surface by the co-catalyst and consequently of the action of the incident photons.[41] This result also supports the hypothesis that the NPs are acting as co-catalysts.

Fig. 6b compares the $H_2$ photocatalytic production of samples before and after thermal treatment, *i.e.,* $Co(OH)_2$ or $Co_3O_4$ NPs as co-catalysts. After thermal treatment, an increase in the crystallinity is expected, and consequently, a decrease in lattice defects, which should, *a priori,* increase the photocatalytic performance. However, the opposite was observed, showing that Co hydroxide displays superior photocatalytic activity as a co-catalyst when introduced to the surface of nanostructured hematite.

As the hydroxide is composed of intercalated Co+ and OH- layers, it can explain the higher $H_2$ evolution rate. The way that the NPs create the reaction sites on the NR surface seems to be controlling the $H_2$ evolution rate. Shimizu *et al.* have reported that the layered tantalates with hydrated interlayer spaces show a higher rate of $H_2$ evolution than that of anhydrous tantalates.[42] Jang *et al.* also reported that $Ni(OH)_2$ was more effective as a co-catalyst than NiO.[40]

The results herein show that hematite nanorings can be used as efficient photocatalysts for $H_2$ generation through water splitting without applying an external bias. In addition, $Co(OH)_2$ nanoparticles, which are composed of intercalated Co+ and OH- layers, proved to be efficient materials as co-catalysts on the surface of the IONRs. Finally, this material showed to be an efficient environmentally friendly catalyst for hydrogen production reaching more than 500 mmol h$^{-1}$ g$^{-1}$, a 4-fold increase with respect to $TiO_2$ nanoparticles (P25). Planned future experiments will probe the electronic structure and band edge energy levels of these hydrothermally prepared IONRs in order to better understand the influence of shape and size on the photocatalytic performance of the materials discussed.

## 4  Conclusions

In summary, IONRs were successfully synthesized through hydrothermal treatment and their surface was coated with $Co(OH)_2$ NPs using chemical precipitation. XANES and XRD results showed that the obtained IONRs are composed of pure hematite phase. Rietveld refinement allowed quantification of the amount of $Co(OH)_2$ NPs deposited on the IONR surface as well as the crystal size of both phases. FESEM and HRTEM images confirmed that the sizes of the $Fe_2O_3$ NRs and $Co(OH)_2$ NPs are approximately 125 and 12 nm respectively. Surprisingly, the as-prepared $Fe_2O_3$ NRs were shown to be active in hydrogen generation by photocatalysis. The observed activity might be attributed to the size and shape properties of the nanostructured hematite due to possible changes in the conduction and valence band positions. By coating the NRs with suitable amounts of $Co(OH)_2$ NPs, which acted as co-catalysts, the photocatalytic activity was further increased. These $Co(OH)_2$ NPs were converted to $Co_3O_4$ NPs after annealing and the photocatalytic activity for hydrogen generation decreased. Both the NP-coated and uncoated NRs displayed superior photo- catalytic activity for hydrogen evolution when compared with $TiO_2$ NPs (P25-Degussa) measured under the same conditions, which proves that these materials are suitable for future studies regarding hydrogen generation.

## Acknowledgements

The authors gratefully acknowledge the Brazilian Synchrotron Light Laboratory (LNLS) for XAFS1 and XAFS2 experimental facilities (proposal XAFS1-12826 and internal research), and LNNano for HRTEM, FESEM and TEM-FEG microscopes (proposals SEM-FEG 13057, TEM-HR 13233, Inspect-13251). The authors also thank the following funding agencies: FAPESP under process no. 2011/17402-9; CNPq under process no. 471220/2010; FAPERGS under process no. 11/2000-4, and ANEEL-CEEE GT under process no. 9945481.

## Notes and references